\documentclass[aps,pre,twocolumn,showpacs,groupedaddress]{revtex4}

\usepackage{graphicx}
\usepackage{amssymb}
\usepackage{wasysym}

\begin{document}

\title{Influence of confinement on granular penetration by impact}
\author{Antoine Seguin, Yann Bertho, and Philippe Gondret}
\affiliation{Univ Paris-Sud, Univ Paris 6, CNRS\\Lab FAST, B\^at.
502, Campus Univ, F-91405 Orsay, France}

\begin{abstract}
We study experimentally the influence of confinement on the penetration depth of impacting spheres into a granular medium contained in a finite cylindrical vessel. The presence of close lateral walls reduces the penetration depth, and the characteristic distance for these lateral wall effects is found to be of the order of one sphere diameter. The influence of the bottom wall is found to have a much shorter range.
\end{abstract}


\maketitle

The motion of a solid sphere falling under its own weight into an unbounded viscous fluid has been known for more than one century to be characterized by the limiting Stokes velocity, and the influence of a confinement by close walls on the sphere velocity is also known for many years both theoretically \cite{Lorentz06, Faxen23, Brenner61} (see Ref.~\cite{Happel86} for a review) and experimentally \cite{MacKay63, Ambari84}. This wall influence, due to the long range of hydrodynamic forces for low Reynolds number flows, results in a larger fluid friction force on the sphere. The velocity of a sedimenting sphere is thus known to be reduced near a wall either parallel or perpendicular to the sphere motion by a correction factor that is linear with $d/l$ at low $d/l$ ($d/l<1$), where $d$ is the sphere diameter and $l$ is the distance from the wall. In all these cases, the well known equations of classical fluid mechanics serve as a reliable support to understand and predict the corresponding motions with a fluid friction always
proportional to velocity at small Reynolds numbers.

Considering now the motion of a falling sphere into granular matter, the existence of a non zero solid friction force at vanishing velocity explains that the sphere driven by its own weight cannot reach a constant limit velocity but stops suddenly at a given depth. As the precise rheology of granular matter is far from being well understood and does not benefit from well-accepted equations despite recent important progress \cite{GdRMiDi04}, the predictions for the stopping distance are far from easy. Many studies have been performed on the penetration depth $\delta$ of an impacting sphere into a granular medium. It emerges from those studies \cite{DeBruyn04, Uehara03, Ambroso05} that $\delta$ follows a power law of the form $\delta /d \propto (\rho /\rho_g)^{\beta}(H/d)^\alpha$, where $\rho_g$ is the density of the grains, $d$ and $\rho$ are respectively the diameter and the density of the sphere and $H$ the total falling distance covered from release to rest. The first power exponent is $\beta \simeq 0.5$ whereas the second, $\alpha$, varies between 0.3 and 0.5 depending mainly on the range of impact velocities and slightly on the packing fraction of the layer \cite{DeBruyn04, Uehara03, Ambroso05}. Indeed, as the packing varies from dense to loose random packing, the packing density does not vary so much but the contact network and force network do vary significantly.

The knowledge and the expression of the forces exerted by the granular medium, and responsible for the sphere deceleration and stop, are still the subject of an intense debate \cite{Hou05, Katsuragi07}, with different terms of frictional and collisional origins. Recently, the effect of a solid wall normal to the sphere motion was shown to account for an exponential increase of the force on the sphere in the close vicinity of the wall \cite{Stone04, Stone04b}.

To our knowledge, no study deals in detail with the influence of close walls on the granular penetration depth, except first results of Ref.~\cite{Goldman08}. In the present paper, after checking the usual ``unbounded" case, we investigate experimentally the influence of the walls, either parallel or perpendicular to the sphere motion, for a sphere dropped onto a granular layer. In some situations, a coupling between the effects of parallel and normal walls may exist, but to understand their respective roles, each is studied separately in this paper.

\section{Experimental setup}

\begin{figure}[b]
\centering
\includegraphics[width=5cm]{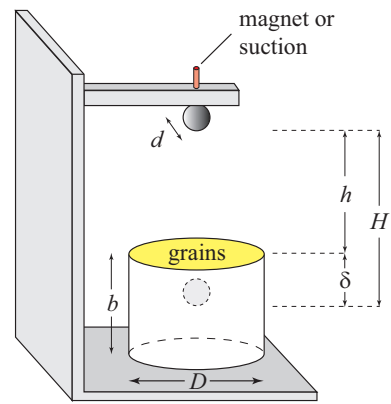}
\caption{Sketch of the experimental setup and notations.}
\label{fig1}
\end{figure}

The experimental setup is sketched in Fig.~\ref{fig1}. The penetration depth $\delta$ is investigated by dropping a spherical projectile of diameter $d$ and density $\rho$ onto a fine granular material in a cylindrical container of diameter $D$ and height $b$.
The granular medium consists of glass beads (density $\rho_g\simeq 2.5\times 10^3$\,kg m$^{-3}$) slightly polydisperse in size, with a diameter range of 300--400\,$\mu$m. Before each drop, the granular medium is prepared by gently stirring the grains with a thin rod. The container is then overfilled and the surface levelled using a straightedge. We have checked that this preparation leads to reproducible results with only small variations. The grain size is much smaller than the falling sphere diameter $d$ so that the granular medium can be considered as a continuum medium.

Different sphere materials and sizes have been used to bring out the influence of the sphere density $\rho$ and of the sphere diameter $d$ on the penetration depth $\delta$. Steel spheres are initially maintained by a magnet at a distance $h$ above the granular surface. Non-metallic spheres are held by creating locally a vacuum at the top of the sphere. Both apparatus allow one to drop the spheres without any initial velocity nor spinning motion. The sphere is released directly above the center of the container and fall along the container axis. The penetration depth $\delta$ is then measured with a thin probe by locating the top of the sphere with a precision better than 1\,mm. The impact speed is tuned by varying the drop height $h$ from 0 to 0.5\,m; the corresponding velocity $v$ at impact, given by $v=\sqrt{2gh}$ where $g$ is the gravitational acceleration, thus ranges from 0 to 3\,m s$^{-1}$. In the following, the results will be analyzed using the total falling distance $H=h+\delta$ (Fig.~\ref{fig1}).

\section{The unbounded case}

\begin{figure}[t]
\centering
\includegraphics[width=8cm]{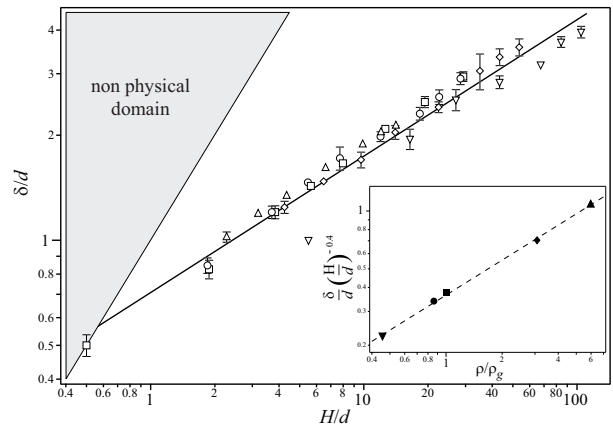}
\caption{Normalized penetration depth $\delta /d$ as a function of the normalized total falling distance $H/d$. Experimental data for steel spheres ($\rho \simeq 7.8\times 10^3$\,kg m$^{-3}$) of different diameters: ($\triangledown$)~$d=5$\,mm, ($\diamond$)~$d=10$\,mm, ($\square$)~$d=19$\,mm, ($\circ$)~$d=20$\,mm and ($\vartriangle$)~$d=40$\,mm. (---) Power law fit $(\delta /d) \propto (H/d)^\alpha$ with $\alpha = 0.4$. The shaded region corresponding to $H<\delta$ is by definition not allowed. Inset: $(\delta /d)(H/d)^{-0.4}$ as a function of the density ratio $\rho /\rho_g$ for spheres of diameter $d=20$\,mm and of different materials: ($\blacktriangledown$)~polyamide, ($\bullet$)~Teflon, ($\blacksquare$)~glass, ($\blacklozenge$)~steel and ($\blacktriangle$)~tungsten carbide. (--~--) Power law fit $(\delta /d)(H/d)^{-0.4} \propto (\rho /\rho_g)^\beta$ with $\beta = 0.61$.}
\label{fig2}
\end{figure}

We first present briefly our results obtained in the usual unconfined case: the container is large enough ($D=190$\,mm $\gg d$) so that the sphere is not affected by the surrounding walls and also high enough ($b=300$\,mm $\gg \delta$) to avoid any bottom wall effect as we shall see in the next sections. The influence of the different parameters on the penetration depth is shown in Fig.~\ref{fig2} for five different sphere materials with a density ranging from $1.14\times 10^3$ (polyamide) to $14.97\times 10^3$\,kg m$^{-3}$ (tungsten carbide) and different sphere diameters $d$ ranging from 5 to 40\,mm. Note that each point corresponds to an average of about ten experiments, and error bars corresponding to the standard deviation are displayed only when larger than the symbol size. The penetration depth is larger for larger falling distance and larger projectile
density. The data are well fitted by power laws of the form
\begin{equation}
\frac{\delta}{d} =A \left (\frac{\rho}{\rho_g}\right )^\beta \left
(\frac{H}{d}\right )^\alpha, \label{eq1}
\end{equation}
with $A =0.37 \pm 0.01$, $\beta = 0.61 \pm 0.02$ and $\alpha = 0.40 \pm 0.04$. The values of $\alpha$ and $\beta$ are in close agreement with the previous studies already mentioned \cite{Uehara03, Ambroso05}.

\section{Influence of lateral confinement}

\begin{figure}[t!]
\centering
\includegraphics[width=8cm]{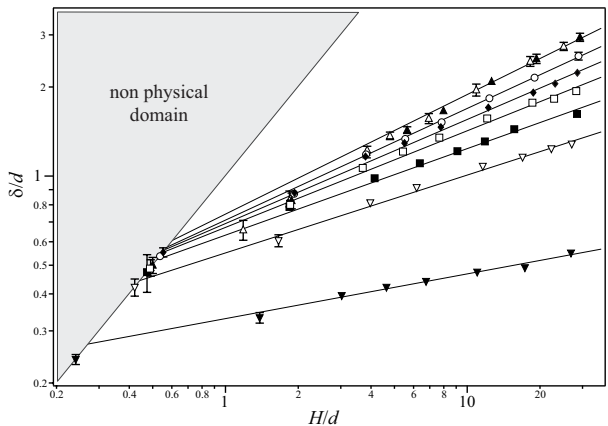}
\caption{Normalized penetration depth $\delta /d$ for a steel sphere ($d=19$\,mm) as a function of the normalized total falling distance $H/d$ for different cylindrical vessel diameters: ($\blacktriangledown$)~$D=24$\,mm, ($\triangledown$)~$D=35$\,mm, ($\blacksquare$)~$D=40$\,mm, ($\square$)~$D=50$\,mm, ($\blacklozenge$)~$D=62$\,mm, ($\circ$)~$D=80$\,mm, ($\blacktriangle$)~$D=128$\,mm, and ($\vartriangle$)~$D=190$\,mm.}
\label{fig3}
\end{figure}

\begin{figure}[t!]
\centering
\includegraphics[width=8cm]{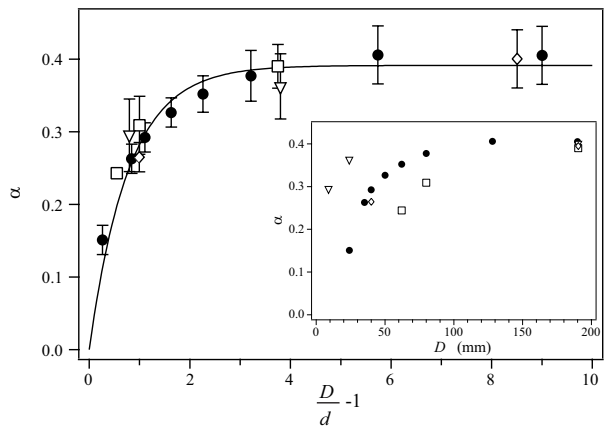}
\caption{Power exponent $\alpha$ of Eq.~(\ref{eq1}) as a function of ($D/d-1$) for steel sphere of diameter ($\triangledown$)~$d=5$\,mm, ($\bullet$)~$d=19$\,mm and ($\square$)~$d=40$\,mm, and ($\diamond$) tungsten carbide sphere of diameter $d=40$\,mm. (---)~Best fit by Eq.~(\ref{eq2}) with $\alpha_\infty \simeq 0.39$ and $\lambda_\alpha\simeq 0.8$. Inset: same data as a function of $D$.}
\label{fig4}
\end{figure}

\begin{figure}[t]
\centering
\includegraphics[width=8cm]{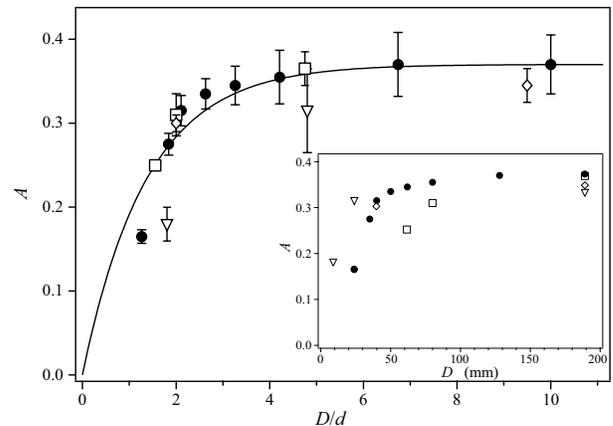}
\caption{Prefactor $A$ of Eq.~(\ref{eq1}) as a function of $D/d$ with the same notations as in Fig.~\ref{fig4}. (---)~Best fit by Eq.~(\ref{eq3}) with $A_\infty \simeq 0.37$ and
$\lambda_A\simeq 0.7$. Inset: same data as a function of $D$.}
\label{fig5}
\end{figure}

We are now interested in the influence of a lateral confinement on the penetration depth, using cylindrical containers of large height ($b=300$\,mm) but with smaller diameters $D$ ranging from 128 down to 24\,mm. The penetration depth $\delta$ is displayed in Fig.~\ref{fig3} as a function of the total falling distance $H$, for a steel projectile of diameter $d=19$\,mm falling in the different vessels. The two data sets corresponding to the two largest containers ($D=128$\,mm and $D=190$\,mm) coincide, suggesting that for large enough vessel diameters the influence of the surrounding walls becomes negligible and the medium can be considered as unbounded in the radial direction. For smaller diameters $D$, data still fall into straight lines in the $\log$--$\log$ plot of Fig.~\ref{fig3} but with smaller slope values suggesting that Eq.~(\ref{eq1}) remains valid with smaller $\alpha$ values. The $A$ and $\alpha$ values extracted from the fit of the data of Fig.~\ref{fig3} by Eq.~(\ref{eq1}) are reported in Figs.~\ref{fig4} and \ref{fig5}.

The inset of Fig.~\ref{fig4} displays $\alpha$ as a function of $D$ for the steel sphere of diameter $d=19$\,mm ($\bullet$). A constant plateau value appears at large enough diameters $D$ ($D \gtrsim 100$\,mm) with the constant value $\alpha \simeq 0.4$. At smaller $D$, $\alpha$ decreases significantly with $D$ and seems to tend towards zero when $D$ approaches the sphere diameter $d=19$\,mm, as no deep penetration (larger than $d/2$) would be possible for $D < d$. Data for different sphere diameters do not collapse in the $\alpha$ vs $D$ plot, but they do collapse in the $\alpha$ vs ($D/d-1$) plot, as shown in Fig.~\ref{fig4}. In this plot, data are well fitted by the exponential law
\begin{equation}
\alpha=\alpha_{\infty}\left[1-\exp\left(-\frac{D-d}{\lambda_\alpha d}\right)\right], \label{eq2}
\end{equation}
where $\alpha_{\infty} \simeq 0.4$ corresponds to the ``infinite" (unbounded) case and $\lambda_\alpha\simeq 0.8$ characterizes the range of wall effects. The characteristic distance $\lambda_\alpha d$ of lateral wall effects is thus found here a little smaller than one sphere diameter, which is rather small. Hence, for $D/d \lesssim 5$, the surrounding walls play a key part and prevent a deep penetration of the projectile. For $D/d \gtrsim 5$, the container has a vanishing influence and the sphere reaches a
limiting penetration depth independent of $D$.

The prefactor $A$ follows the same kind of decrease for decreasing container diameters $D$ (see Fig.~\ref{fig5}), but we believe that the relevant parameter for the lateral confinement is here $D/d$ rather than ($D/d-1$), as $A$ must vanish for vanishing $D/d$. The $A$ variation is quite well fitted by the exponential law
\begin{equation}
A=A_{\infty}\left[1-\exp\left(-\frac{D}{\lambda_A d}\right)\right],
\label{eq3}
\end{equation}
where $A_\infty \simeq 0.37$ corresponds to the unbounded case and $\lambda_A \simeq 0.7$ characterizes the range of wall effects. Note that the $\lambda_\alpha$ and $\lambda_A$ values are about the same as these two parameters characterize the same wall effect.

Besides, the power exponent $\beta$ for the density ratio [Eq.~(\ref{eq1})] does not depend on the lateral confinement but may be considered as constant, as data points for tungsten carbide ($\diamond$) twice denser than steel fall on the same curve in both Figs. 4 and 5.

The reduction of the penetration depth by a lateral confinement is clearly due to an enhanced blocking effect by the walls (less radial dilatancy). The precise mechanisms responsible for this reduction remain to be understood. For instance, the non-linear ``pressure" evolution inside the granular packing due to the well-known Janssen effect, and valid both in static and dynamic situations \cite{Bertho03}, fails in explaining an enhanced force on the sphere in the confined case.

\section{Influence of normal confinement}

\begin{figure}[t]
\centering
\includegraphics[width=8cm]{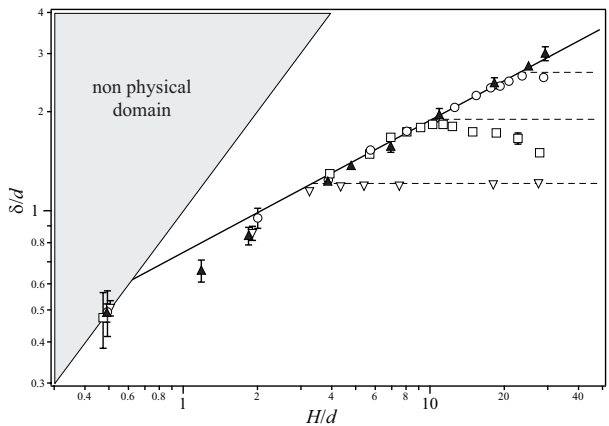}
\caption{Normalized penetration depth $\delta /d$ of a steel sphere ($d=19$\,mm) as a function of the normalized total drop height $H/d$, for different heights $b$ of the granular layer: ($\triangledown$)~$b/d=1.2$, ($\square$)~$b/d=1.9$, ($\circ$)~$b/d=2.6$, ($\blacktriangle$)~$b/d=15.8$. Each horizontal dashed line corresponds to each initial height $b$ of the granular layer. (---)~Power law fit $(\delta /d)\propto (H/d)^\alpha$ with $\alpha = 0.4$, for the unbounded case ($b/d=15.8$).} \label{fig6}
\end{figure}

Let us now discuss the influence of a normal confinement by a close bottom wall. We investigate this effect by varying the thickness $b$ of the granular layer contained in a vessel of large enough diameter $D$ to avoid the lateral wall effects discussed before. As we are interested in the sphere penetration inside the granular material, we restrict however our study to thick enough layers with $b>d$. The normalized penetration depth $\delta /d$ of the steel sphere of diameter $d=19$\,mm is displayed in Fig.~\ref{fig6} as a function of the normalized total drop distance $H/d$ for three layer heights $b$ ranging from 23 to 50\,mm ($1.2\leq b/d\leq 2.6$) together with the unbounded case $b=300$\,mm ($b/d=15.8$).
The penetration depth remains unchanged by the presence of the bottom wall until the projectile approaches at a very short distance from the wall: $\delta$ follows the unbounded curve until the sphere impacts the bottom wall indicated by the dashed lines in Fig.~\ref{fig6}. This very short range effect is consistent with recent measurements of the force on a flat plate approaching in a \emph{quasi-static} way the solid bottom boundary of a granular sample, indicating that the penetration resistance increases exponentially near the boundary \cite{Stone04, Stone04b}. At a high enough falling height, the sphere impacts the wall (a characteristic shock sound can be heard) and bounces back leaving a crater below. For thin enough layers as already studied in Ref.~\cite{Boudet06}, the opened crater does not get enough filled by inwards avalanches so that no grains are at the center and the final rest position of the sphere corresponds to $\delta = b$ as the sphere comes to rest on the bottom ``dry" wall [see ($\triangledown$) in Fig.~\ref{fig6}]. For thicker granular layers, inwards avalanches are large enough to bring back grains down to the crater center so that the sphere comes back at rest at $\delta <b$ [see ($\square$) in Fig.~\ref{fig6}]. In that case, the larger is the falling height, the longer is the rebound time and thus the time for avalanche, so that the thicker is the layer of back grains at the crater center: this explains the decreasing value of $\delta$ for an increasing falling height in this part of the curve. Data for $b/d=2.6$ ($\circ$) would also display such a behavior for larger falling heights.

\ \\

\section{Conclusion}

We have measured the penetration depth $\delta$ of spheres of diameter $d$ impacting a granular medium contained in a cylindrical vessel of diameter $D$. This penetration depth may be strongly affected by the presence of surrounding walls. The presence of a bottom wall perpendicular to the sphere motion has a very short range influence on $\delta$ as the sphere is only affected when the distance from the wall has reached a few grain diameters. On the other hand, the presence of lateral walls parallel to the sphere motion has a larger range influence with a typical characteristic distance $\lambda \simeq d$. For $D/d\lesssim 5$, the walls have a strong influence and reduce significantly the penetration depth. For $D/d\gtrsim 5$, they have a vanishing influence. The two (bottom and lateral) wall effects have been studied independently but it could be a hard coupling between them for geometrically strongly hindered grain motions.

\section*{Acknowledgements}

We wish to thank M. Salloum for preliminary experiments and J.-P. Hulin and M. Rabaud for a thoughtful reading of the manuscript.

\bibliography{../../../biblio/biblio}
\end{document}